\def\be{\begin{equation}}
\def\ee{\end{equation}}
\def\ba{\begin{array}}
\def\ea{\end{array}}

\def\L{\Lambda}
\def\l{\lambda}

\documentclass[sn-mathphys]{sn-jnl}% Math and Physical Sciences Reference Style
\usepackage{amsmath,amssymb,amsfonts}%
\usepackage{mathrsfs}
\usepackage[OT2,OT1]{fontenc}
\usepackage{hyperref}
\usepackage[hyphenbreaks]{breakurl}
\usepackage[T1]{fontenc}
\jyear{2021}%

\theoremstyle{thmstyleone}%
\newtheorem{theorem}{Theorem}

\theoremstyle{thmstyletwo}%

\theoremstyle{thmstylethree}%

\begin{document}

\title[Article Title]{Optimizing incompatible triple quantum measurements}

\author*[1]{\fnm{Hui-Hui}\sur{ Qin}}\email{qinhh@hdu.edu.cn}

\author[2,3]{\fnm{Shao-Ming }\sur{ Fei}}\email{feishm@cnu.edu.cn}
\equalcont{These authors contributed equally to this work.}

\affil*[1]{\orgdiv{School of Sciences}, \orgname{Hangzhou Dianzi University}, \orgaddress{\street{} \city{Hangzhou}, \postcode{310018}, \state{} \country{China}}}

\affil[2]{\orgdiv{Max-Planck-Institute for Mathematics in the Sciences},\orgname{}\orgaddress{\street{} \city{Leipzig}, \postcode{04103}, \state{} \country{Germany}}}

\affil[3]{\orgdiv{School of Mathematical Sciences},\orgname{Capital Normal University}, \orgaddress{\street{}\city{Beijing}, \postcode{100048},\state{} \country{China}}}

\abstract{We investigate the optimal approximation to triple incompatible quantum measurements within the framework of statistical distance and joint measurability. According to the lower bound of the uncertainty inequality presented in [Physical Review A 99, 312107 (2019)], we give the analytical expressions of the optimal jointly measurable approximation to two kinds of triple incompatible unbiased qubit measurements. We also obtain the corresponding states which give the minimal approximation errors in measuring process. The results give rise to plausible experimental verifications of such statistical distance based uncertainty relations.}

\keywords{Joint measurability, Uncertainty relations involving triple incompatible measurements, The optimal approximations, Fermat-Torricelli point}



\maketitle

\maketitle
\section{Introduction}\label{sec1}
Uncertainty relations reveal a part of the essence of quantum physics. Since the Heisenberg's uncertainty relation of error-disturbance \cite{heisenberg} for measurements of incompatible observables, there has been a series of researches on subject of uncertainty relations \cite{robertson29,weigert,chenprl,qinhh,maccone,
pati,chensum,maassen,berta,coles,fan,xiaoyl,puchala,
friedland,tli,xiaoy2,ozawa03,ozawa04,ff,busch07,busch13prl,
busch14pra,busch14rmp,wenma,buscemi,sulyok,yuarxiv,
svetlichny,andreas,alberto,qinhhpra,lars,stephanie,spiros,kamran,maoyali}. In addition to illustrating the impossibility of simultaneously determining the definite values of incompatible observables, the uncertainty inequalities also indicate the minimal amount of errors produced in the measuring process. Because of the inevitable errors during the measuring process it is important to investigate measurement schemes which produce less measurement errors. Based on different measurement schemes, different types of uncertainty relations involving different error-disturbance in the measuring process are widely investigated. In \cite{robertson29,weigert,chenprl,qinhh,maccone,pati,
chensum} the measurement errors are defined by the derivations of observables, while in \cite{maassen,berta,coles,fan,xiaoyl,puchala,friedland,
tli,xiaoy2} the measurement errors are described by entropies. There are also schemes in which the measurement errors stem from the difference between the target observables and the observables measured practically \cite{ozawa03,ozawa04,busch07,busch13prl,busch14pra,
busch14rmp,wenma,yuarxiv,qinhhpra,maoyali,kamran}. A typical case in this scheme is to use the compatible observables $C$ and $D$ to approximate the target observables $A$ and $B$, respectively \cite{busch07,busch13prl,busch14pra}. The total approximation error-disturbance is constrained by a measure of the degree of incompatibility of the target observables $A$ and $B$:
\be\label{eqbusch}
\Delta(C,A)^2+\Delta(D,B)^2\geq (incompatibility~of~A~and~B),
\ee
where the state-independent error-disturbance $\Delta(A,C)^2~(\Delta(B,D)^2)$ is defined as the Wasserstein distance (of order 2) of probability distributions between the positive operator valued measures (POVMs) $A$ and $C$ ($B$ and $D$), and the incompatibility of $A$ and $B$ in the right-hand side of (\ref{eqbusch}) is defined by the non-jointly measurability of $A$ and $B$. In \cite{wenma}, the authors gave the expressions of the optimal compatible observables $C$ and $D$ approximating the incompatible unbiased qubit measurements $A$ and $B$ respectively, and verified experimentally the uncertainty relation $(\ref{eqbusch})$ by measuring the corresponding optimal compatible observables $C$ and $D$ for given incompatible observables $A$ and $B$.

For triple incompatible target measurements $\{M^i\}^3_{i=1}$ it is rational to generalize the approximation scheme above by measuring triple compatible measurements $\{N^i\}^3_{i=1}$ instead.
In \cite{qinhhpra} by defining similar error-disturbance $\Delta(M^i,N^i)^2$ between the measurements $M^i$ and $N^i$ respectively, we obtained an uncertainty relation,
\be\label{qhh}
\sum^3_{i=1}\Delta(M^i,N^i)^2\geq~ (incompatibility ~~of~~\{M^i\}^3_{i=1}),
\ee
where the right-hand side is a quantity which measures the degree of the incompatibility of $\{M^i\}^3_{i=1}$.

In this work we investigate the optimal approximation to triple incompatible target observables. According to different classes of triple incompatible unbiased qubit measurements, we obtain the analytical expressions of the optimal observables in categories. We also obtain the corresponding quantum states which give rise to the minimal approximation error-disturbance. These results are of instructive significance for experimental implementations.

\section{The optimal approximation to triple incompatible unbiased qubit measurements}\label{sec3}

\subsection{The uncertainty relation involving triple incompatible unbiased qubit measurements}
Let $M$ and $N$ be two unbiased qubit POVMs with measurement operators $M_{\pm}=\frac{1}{2}
(\mathbf{1}\pm\vec{m}\cdot\vec{\sigma})$ and $N_{\pm}=\frac{1}{2}(\mathbf{1}\pm\vec{n}\cdot\vec{\sigma})$, respectively,
where $\vec{\sigma}$ is the vector given by the standard Pauli matrices, $\mathbf{1}$ is the $2\times 2$ identity matrix, $\vec{m}$ and $\vec{n}$ are three dimensional real vectors (Bloch vectors) such that $\|\vec{m}\|\leq1$ and
$\|\vec{n}\|\leq 1$. The Wasserstein distance between $M$ and $N$ with respect to the qubit state
$\rho=(\mathbf{1}+\vec{r}\cdot\vec{\sigma})/{2}$ ( $\|\vec{r}\|\leq 1$) is given by $\Delta_{\rho}(M,N)^2=\sum_{i=\pm}Tr\rho(M_i-N_i)
=2\|\vec{r}\cdot(\vec{m}-\vec{n})\|$.
In the scheme of measuring triple incompatible qubit observables,
the three POVMs $\{M^i\}^3_{i=1}$ given by the vectors $\{\vec{m}_i\}^3_{i=1}$ are approximated by three jointly measurable POVMs $\{N^i\}^3_{i=1}$ given by the vectors $\{\vec{n}_i\}^3_{i=1}$, respectively.
The three POVMs $\{N^i\}^3_{i=1}$ are jointly measurable if and only if $\{\vec{n}_i\}^3_{i=1}$ satisfy \cite{yuarxiv,qinhhpra},
\be\label{yu2}
\sum^4_{k=1}\|\vec{q}_F-\vec{q}_k\|\leq4,
\ee
where $\vec{q}_{1}=\vec{n}_{123}$, $\vec{q}_{2}=\vec{n}_{1}-\vec{n}_{23}$, $\vec{q}_{3}=\vec{n}_{2}-\vec{n}_{13}$,
$\vec{q}_{4}=\vec{n}_{3}-\vec{n}_{12}$, with $\vec{n}_{123}=\vec{n}_1+\vec{n}_2+\vec{n}_3$, $\vec{n}_{ij}=\vec{n}_i+\vec{n}_j$ and $\vec{q}_F$ the Fermat-Torricelli point \cite{boltyanski} of the four vectors $\{\vec{q}_k\}^4_{k=1}$.

For any given triple incompatible unbiased qubit measurements $\{M^i\}^3_{i=1}$, we have obtained in \cite{qinhhpra} that the total error-disturbance satisfies the following inequality,
\be\label{ur3}
\sum^3_{i=1}\Delta(M^i,N^i)^2
\hat{=}\max_{\rho}\sum^3_{i=1}
\Delta_{\rho}(M^i,N^i)^2
\geq\frac{1}{2}[\sum^{4}_{k=1}\|\vec{p}_{k}-\vec{p}_{F}\|-4],
\ee
where $\{N^i\}^3_{i=1}$ are triple compatible unbiased qubit measurements, the definition of $\vec{p}_k$s is similar to the definition of $\vec{q}_k$s in (\ref{yu2}) and $\vec{p}_F$ is the Fermat-Torricelli point of $\{\vec{p}_{k}\}^4_{k=1}$. Accounting to the possibility of imperfect preparation of states to be measured, we consider the worst approximation error-disturbance over all quantum states which corresponds to the maximum over all quantum states $\rho$ in (\ref{ur3}).

The lower bound of the inequality (\ref{ur3}) is complicated to be evaluated for general $\{\vec{m}_i\}^3_{i=1}$, since the Fermat-Torricelli point of $\{\vec{p}_{k}\}^4_{k=1}$ given by $\{\vec{m}_i\}^3_{i=1}$ is difficult to analyze. The lower bound of the inequality (\ref{ur3}) is attained if and only if the following conditions are satisfied:

$\romannumeral 1)$. $\vec{p}_k$, $\vec{q}_k$ and $\vec{p}_F$ are collinear for $k=1,2,3,4$;

$\romannumeral 2)$. $\|\vec{p}_k-\vec{q}_k\|=\|\vec{p}_l-\vec{q}_l\|$, $k\neq l$, $k,l=1,2,3,4$;

$\romannumeral 3)$. $\vec{p}_F=\vec{q}_F$;

$\romannumeral 4)$. $\sum^4_{k=1}\|\vec{q}_k-\vec{q}_F\|=4$.

\noindent These conditions are highly dependent on the Fermat-Torricelli point $\vec{p}_F$. But for any given four vertices $\{\vec{p}_k\}^4_{k=1}$ there is no analytical solution of the corresponding Fermat-Torricelli point. According to \cite{boltyanski}, the Fermat-Torricelli point $\vec{p}_F$ of $\{\vec{p}_k\}^4_{k=1}$ satisfies that
\be\label{ftp}
\sum^4_{k=1}\frac{\vec{p}_k-\vec{p}_F}
{\|\vec{p}_k-\vec{p}_F\|}=\vec{0}.
\ee
In the following we investigate the solutions of the Fermat-Torricelli point of $\{\vec{p}_k\}^4_{k=1}$. And then we study the optimal approximation to the triple incompatible unbiased qubit POVMs while the Fermat-Torricelli point of $\{\vec{p}_k\}^4_{k=1}$ can be solved analytically. For the reason that the solutions of Fermat-Torricelli point of $\{\vec{p}_k\}^4_{k=1}$ depends on the relationship between Bloch vectors $\{\vec{m}_i\}^3_{i=1}$ of $\{M^i\}^3_{i=1}$, we study in categories the optimal approximation according to the relationship among $\{\vec{m}_i\}^3_{i=1}$.

\subsection{The optimal approximation when $\vec{m}_3$ is perpendicular to the vectors $\vec{m}_1$ and $\vec{m}_2$}
When the Bloch vector $\vec{m}_3$ is perpendicular to the Bloch vectors $\vec{m}_1$ and $\vec{m}_2$ respectively, $\{\vec{p}_k\}^4_{k=1}$
constitute a tetrahedron. According to (\ref{ftp}) we obtain that the Fermat-Torricelli point $\vec{p}_F$ of $\{\vec{p}_k\}^4_{k=1}$ lies on the line going through the point origin and the point $\vec{m}_3$ and has form
\be\label{ftp1}
\vec{p}_F=\frac{\|\vec{m}_{1-2}\|-\|\vec{m}_{12}\|}
{\|\vec{m}_{1-2}\|+\|\vec{m}_{12}\|}\vec{m}_3,
\ee
where $\vec{m}_{ij}=\vec{m}_i+\vec{m}_j$ and $\vec{m}_{i-j}=\vec{m}_i-\vec{m}_j$ $(i,j=1,2)$.

\begin{figure}[H]
\setlength{\abovecaptionskip}{0.cm}
\setlength{\belowcaptionskip}{-0.cm}
  \centering
  \includegraphics[height=4cm]{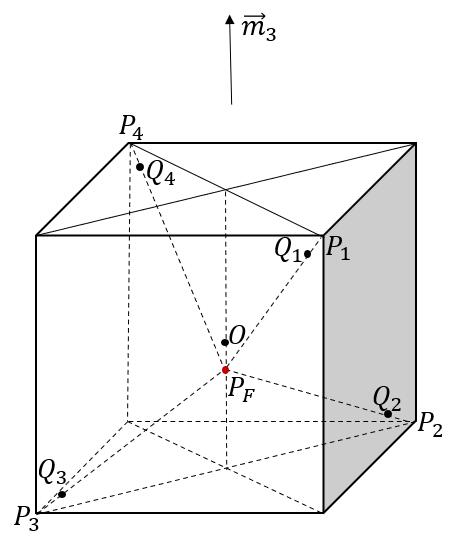}\\
  \caption{The vertices $\{\vec{p}_k\}^4_{k=1}~(\{P_k\}^4_{k=1})$
   constitute a tetrahedron and the Fermat-Torricelli vector $\vec{p}_F$ is parallel to $\vec{m}_3$. The optimal POVMs given by $\{\vec{n}_i\}^3_{i=1}$ can be obtained from $\{\vec{q}_k\}^4_{k=1}$~( $\{Q_k\}^4_{k=1}$) satisfying the conditions $\expandafter{\romannumeral1})-
  \expandafter{\romannumeral4})$.
  }\label{111}
\end{figure}

As is shown in Fig. \ref{111}, we need to analyze the solution of vertices $\{\vec{q}_k\}^4_{k=1}$ constituted by the optimal $\{\vec{n}_i\}^3_{i=1}$, i.e., the corresponding $\{N^i\}^3_{i=1}$ are the optimal approximation. The first condition $\expandafter{\romannumeral1})$ means that $\vec{q}_k~(Q_k)$ lies on the segment going through $\vec{p}_F~(P_F)$ and $\vec{p}_k~(P_k)$~$(k=1,2,3,4)$. The second condition $\expandafter{\romannumeral2})$ means that $\|\vec{p}_k-\vec{q}_k\|=\|\vec{p}_l-\vec{q}_l\|~
(\|P_kQ_k\|=\|P_lQ_l\|)$~$(k,l=1,2,3,4)$. If the third condition $\expandafter{\romannumeral3})$ holds, then the vertices $\{\vec{q}_k\}^4_{k=1}$ satisfying $\sum^4_{k=1}\|\vec{q}_k-\vec{p}_F\|=4$
are the solution of the problem. Apparently $\vec{p}_F$ satisfies that
$\sum^4_{k=1}\frac{\vec{q}_k-\vec{p}_F}
{\|\vec{q}_k-\vec{p}_F\|}=\sum^4_{k=1}\frac{\vec{p}_k-\vec{p}_F}
{\|\vec{p}_k-\vec{p}_F\|}=\vec{0}$,
while $\vec{q}_k$ lies on the segment going through $\vec{p}_F$ and $\vec{p}_k$ $(k=1,2,3,4)$. It means that the Fermat-Torricelli point of $\{\vec{q}_k\}^4_{k=1}$ coincides with $\vec{p}_F$.  Then for these $\{\vec{q}_k\}^4_{k=1}$ the third condition holds as well. Thus the optimal POVMs $\{N^i\}^3_{i=1}$ given by $\{\vec{n}_i\}^3_{i=1}$ can be obtained from $\{\vec{q}_k\}^4_{k=1}$. And from the geometrical position of $\{\vec{q}_k\}^4_{k=1}$ we have that the optimal approximation $\{\vec{n}_i\}^3_{i=1}$ is unique.
While $\vec{m}_3$ is parallel to $\vec{p}_F$ and perpendicular to $\vec{m}_1, \vec{m}_2$, we have that
\be
\left\{
\begin{aligned}
&\|\vec{p}_1-\vec{p}_F\|=\|\vec{m}_{12}+(\vec{m}_3-\vec{p}_F)\|=
\|-\vec{m}_{12}+(\vec{m}_3-\vec{p}_F)\|=\|\vec{p}_4-\vec{p}_F\|,\\
&\|\vec{p}_2-\vec{p}_F\|=\|\vec{m}_{1-2}-(\vec{m}_3+\vec{p}_F)\|=
\|\vec{m}_{2-1}-(\vec{m}_3+\vec{p}_F)\|=\|\vec{p}_3-\vec{p}_F\|.
\end{aligned}
\right.
\ee
In order to calculate $\{\vec{q}_k\}^4_{k=1}$ we assume that $\|\vec{p}_k-\vec{q}_k\|=\mu\|\vec{p}_1-\vec{p}_F\|$ for $k=1,4$ and $\|\vec{p}_l-\vec{q}_l\|=\nu\|\vec{p}_2-\vec{p}_F\|$ for $l=2,3$, where $\mu,\nu\in[0,1]$. Then from the conditions $\expandafter{\romannumeral2})$ and $\expandafter{\romannumeral4})$ we obtain
\be\label{munu}
\left\{
\begin{aligned}
&\mu\|\vec{p}_1-\vec{p}_F\|=\|\vec{p}_k-\vec{q}_k\|=
\|\vec{p}_l-\vec{q}_l\|=\nu\|\vec{p}_2-\vec{p}_F\|~(k=1,4,l=2,3),\\
&(1-\mu)\|\vec{p}_1-\vec{p}_F\|+(1-\nu)\|\vec{p}_2-\vec{p}_F\|=2.
\end{aligned}
\right.
\ee
From (\ref{munu}) above we obtain the solution of $\mu,\nu$,
\be\label{mu}
\begin{aligned}
&\mu=\frac{1}{2}(1-\frac{1}{\|\vec{p}_F-\vec{p}_1\|}
+\frac{\|\vec{p}_F-\vec{p}_2\|}{\|\vec{p}_F-\vec{p}_1\|}),\\
&\nu=\frac{1}{2}(1-\frac{1}{\|\vec{p}_F-\vec{p}_2\|}
+\frac{\|\vec{p}_F-\vec{p}_1\|}{\|\vec{p}_F-\vec{p}_2\|}).
\end{aligned}
\ee
Therefore, we have the following conclusion:

\begin{theorem}[]\label{thm1}
The optimal jointly measurable POVMs $\{N^i\}^3_{i=1}$ are given by
\be
\left\{
\begin{aligned}
&\vec{n}_1=
\frac{1}{2}\Big[(2-\mu-\nu)\vec{m}_1+(\nu-\mu)\vec{m}_2
+(\nu-\mu)\vec{m}_3+(\mu+\nu)\vec{p}_F\Big],\\
&\vec{n}_2=
\frac{1}{2}\Big[(\nu-\mu)\vec{m}_1+(2-\mu-\nu)\vec{m}_2
+(\nu-\mu)\vec{m}_3+(\mu+\nu)\vec{p}_F\Big],\\
&\vec{n}_3=(1-\mu)\vec{m}_3+\mu\vec{p}_F,
\end{aligned}
\right.
\ee
where $\mu$ and $\nu$ are given by (\ref{mu}).
\end{theorem}

{\sf [Proof]} According to the analysis above we have that $\vec{q}_k$ lies on the segment going through $\vec{p}_k,\vec{p}_F$ and $\|\vec{p}_k-\vec{q}_k\|=\mu\|\vec{p}_k-\vec{p}_F\|$ for $k=1,4$. It means that $\vec{p}_k-\vec{q}_k=\mu(\vec{p}_k-\vec{p}_F)$,~$k=1,4$.
At the same time $\vec{q}_l$ lies on the segment going through $\vec{p}_l,\vec{p}_F$ and  $\|\vec{p}_l-\vec{q}_l\|=\nu\|\vec{p}_2-\vec{p}_F\|$ for $l=2,3$, which means that $\vec{p}_l-\vec{q}_l=\nu(\vec{p}_l-\vec{p}_F)$,~$l=2,3$. This leads to
\be\label{q1234}
\begin{aligned}
&\vec{q}_1=(1-\mu)\vec{p}_1+\mu\vec{p}_F,\\
&\vec{q}_2=(1-\nu)\vec{p}_2+\nu\vec{p}_F,\\
&\vec{q}_3=(1-\nu)\vec{p}_3+\nu\vec{p}_F,\\
&\vec{q}_4=(1-\mu)\vec{p}_4+\mu\vec{p}_F.
\end{aligned}
\ee
Thus we obtain that the optimal $\{\vec{n}_i\}^3_{i=1}$ has form
\be
\left\{
\begin{aligned}
&\vec{n}_1=\frac{1}{2}(\vec{q}_1+\vec{q}_2)=\frac{1}{2}\Big[(2-\mu-\nu)\vec{m}_1+(\nu-\mu)\vec{m}_2
+(\nu-\mu)\vec{m}_3+(\mu+\nu)\vec{p}_F\Big],\\
&\vec{n}_2=\frac{1}{2}(\vec{q}_1+\vec{q}_3)=
\frac{1}{2}\Big[(\nu-\mu)\vec{m}_1+(2-\mu-\nu)\vec{m}_2
+(\nu-\mu)\vec{m}_3+(\mu+\nu)\vec{p}_F\Big],\\
&\vec{n}_3=\frac{1}{2}(\vec{q}_1+\vec{q}_4)=(1-\mu)\vec{m}_3+\mu\vec{p}_F,
\end{aligned}
\right.
\ee
where $\mu$ and $\nu$ are given by (\ref{mu}). \qed

The maximization in (\ref{ur3}) over the measured state $\rho$ guarantees that the minimal statistical distance between POVMs $\{M^i\}^3_{i=1}$ and the jointly measurable POVMs $\{N^i\}^3_{i=1}$ would not change due to the improper preparing of measured states. In \cite{qinhhpra} we have shown that the optimal $\rho$ has Bloch vector
$\vec{r}_o=\frac{\vec{s}}{\|\vec{s}\|}$, where $\vec{s}$ satisfies that $\|\vec{s}\|=\max_k\|\vec{p}_k-\vec{q}_k\|$. According to the second condition $\expandafter{\romannumeral2})$, we have
\be\label{vecr}
\vec{r}_o=\frac{\vec{p}_k-\vec{q}_k}
{\|\vec{p}_k-\vec{p}_F\|}=
\frac{\vec{p}_k-\vec{p}_F}{\|\vec{p}_k-\vec{p}_F\|},~ ~ k=1,2,3,4.
\ee
Substituting $\vec{p}_ks$ and $\vec{p}_F$ into (\ref{vecr}) we obtain
\be
\vec{r}_o\in\{\frac{\vec{m}_{12}+\frac{2x}{x+y}\vec{m}_3}
{\|\vec{m}_{12}+\frac{2x}{x+y}\vec{m}_3\|},
\frac{(\vec{m}_1-\vec{m}_2)-\frac{2y}{x+y}\vec{m}_3}
{\|(\vec{m}_1-\vec{m}_2)-\frac{2y}{x+y}\vec{m}_3\|}\},
\ee
where $x=\|\vec{m}_{12}\|$ and $y=\|\vec{m}_{1-2}\|$.

In order to illustrate the Theorem 1, we calculate the optimal approximation for triple sharp qubit observables, i.e., the projective measurements. Without loss of generality, we assume that $\vec{m}_3=(0,0,1)$ and parameterize $\vec{m}_1,\,\vec{m}_2$ as:
$\vec{m}_1=(-\sin\alpha,\cos\alpha,0)$, $\vec{m}_2=(\sin\beta,\cos\beta,0)$.
Then $\vec{p}_1=\vec{m}_{123}=(-\sin\alpha+\sin\beta,\cos\alpha+\cos\beta,1)$,
$\vec{p}_2=\vec{m}_{1}-\vec{m}_{23}=(-(\sin\alpha+\sin\beta),\cos\alpha-\cos\beta,-1)$,
$\vec{p}_3=\vec{m}_{2}-\vec{m}_{13}=(\sin\alpha+\sin\beta,-\cos\alpha+\cos\beta,-1)$ and
$\vec{p}_4=\vec{m}_{3}-\vec{m}_{12}=(\sin\alpha-\sin\beta,-(\cos\alpha+\cos\beta),1)$.
By substituting $\{\vec{m}_i\}^3_{i=1}$ into the equation (\ref{ftp1}) we obtain that
\be\label{ftp2}
\vec{p}_F=\frac{\|\vec{m}_{1-2}\|-\|\vec{m}_{12}\|}
{\|\vec{m}_{1-2}\|+\|\vec{m}_{12}\|}\vec{m}_3
=(0,0,-\frac{\cos(\alpha+\beta)}
{1+\mid\sin(\alpha+\beta)\mid}).
\ee
We can then calculate the form of parameter $\mu,\nu$ in $(\ref{mu})$
$$
\left\{
\begin{aligned}
&\mu=\frac{1}{2}\Big[1+\frac{\mid\sin(\alpha+\beta)\mid}{1+\cos(\alpha+\beta)}-
\sqrt{\frac{2(1+\mid\sin(\alpha+\beta)\mid)}{(1+\cos(\alpha+\beta))(2+\mid\sin(\alpha+\beta)\mid)}}~\Big],\\
&\nu=\frac{1}{2}\Big[1+\frac{1+\cos(\alpha+\beta)}{\mid\sin(\alpha+\beta)\mid}-
\sqrt{\frac{2(1+\mid\sin(\alpha+\beta)\mid)}{(1-\cos(\alpha+\beta))(2+\mid\sin(\alpha+\beta)\mid)}}~\Big].
\end{aligned}
\right.
$$

More specifically, for three Pauli observables $\sigma_x,\sigma_y,\sigma_z$, the corresponding Bloch vectors $\{\vec{m}_i\}^3_{i=1}$ are unit and perpendicular to each other. We have in this case $\vec{p}_F=(0,0,0)$ and $\mu=\nu=1-\frac{1}{\sqrt{3}}$. Then the optimal  jointly measurable approximation $\{N^i\}^3_{i=1}$ is given by $\vec{n}_i=\frac{1}{\sqrt{3}}\vec{m}_i$ $(i=1,2,3)$. The corresponding optimal state $\rho=\frac{1}{2}(I+\vec{r}\cdot\vec{\sigma})$ can be given by
$\vec{r}=\vec{r}_o=\frac{1}{\sqrt{3}}(1,1,1)$, $\frac{1}{\sqrt{3}}(1,-1,-1)$, $\frac{1}{\sqrt{3}}(-1,1,-1)$ or $\frac{1}{\sqrt{3}}(-1,-1,1)$, as shown in \cite{qinhhpra}.

\subsection{The optimal approximation when $\vec{m}_3$ is coplanar to the vectors $\vec{m}_1$ and $\vec{m}_2$}

When the measurement $M^3$'s Bloch vector $\vec{m}_3$ is coplanar to the Bloch vectors $\vec{m}_1$ and $\vec{m}_2$ of $M^1$ and $M^2$ respectively, $\{\vec{p}_k\}^4_{k=1}$ lie on the same plane of $\{\vec{m}_i\}^3_{i=1}$. Without loss of generality, we assume that $\vec{m}_3=k_1\vec{m}_1+k_2\vec{m}_2$~
$(\|\vec{m}_i\|\leq1,i=1,2,3)$.
Obviously the Fermat-Torricelli point $\vec{p}_F$ of $\{\vec{p}_k\}^4_{k=1}$ lies on the plane. According to \cite{boltyanski} we obtain that the solution of the Fermat-Torricelli point $\vec{p}_F$ depends on the convexity of the quadrilateral constituted by $\{\vec{p}_k\}^4_{k=1}$. We study the issues in terms of the following two cases.

\begin{figure}[H]
\setlength{\abovecaptionskip}{0.cm}
\setlength{\belowcaptionskip}{-0.cm}
  \centering
  \includegraphics[width=0.3\textwidth,height=4cm]{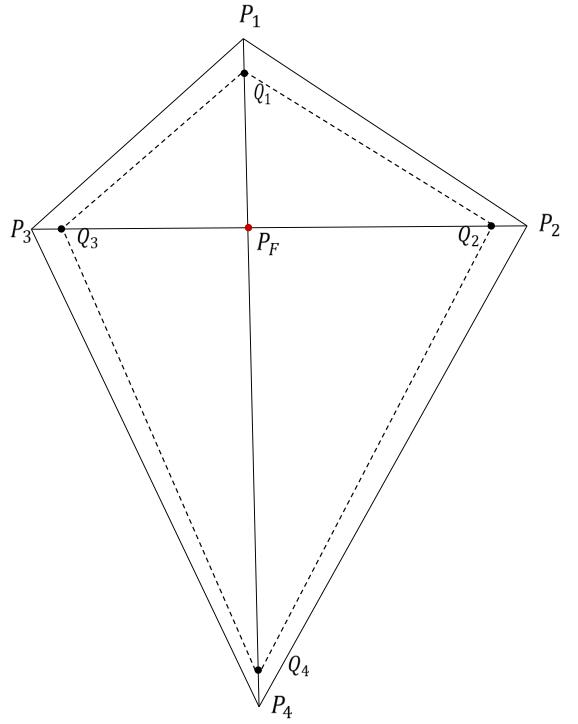}
  \caption{The vertices $\{P_k\}^4_{k=1}$ associated with $\{\vec{p}_k\}^4_{k=1}$ constitute a convex quadrilateral. The Fermat-Torricelli point $\vec{p}_F$ is on the intersection of the diagonals. The optimal approximation of jointly measurable POVMs $\{N^i\}^3_{i=1}$ given by $\{\vec{n}_i\}^3_{i=1}$ is obtained from $\{\vec{q}_k\}^4_{k=1}~(\{Q_k\}^4_{k=1})$ on the plane with $\{\vec{q}_k\}^4_{k=1}$ satisfying the conditions
  $\expandafter{\romannumeral1})-\expandafter{\romannumeral4})$.
  %~
  %\vec{p}_k,\vec{q}_k,\vec{p}_F~are ~collinear,~~for~k=
  %1,2,3,4$;
  %$\expandafter{\romannumeral2}).~
  %|\vec{p}_k-\vec{q}_k|=
  %|\vec{p}_l-\vec{q}_l|~(k\neq l;k,l=1,2,3,4)$;
  %$\expandafter{\romannumeral3}).~ \vec{p}_F=\vec{q}_F$;
  %$\expandafter{\romannumeral4}). ~ \sum^4_{k=1}|\vec{q}_k-\vec{p}_F|=4$.
  }\label{1111}
\end{figure}

First we consider the situation when $\mid k_1\mid+\mid k_2\mid<1$. In this situation $\{\vec{p}_k\}^4_{k=1}$ constitute a convex quadrilateral. We just need to analyze the case when $k_1,\,k_2>0$ and $\mid k_1\mid+ \mid k_2\mid<1$, as shown in Fig. \ref{1111}. Other cases correspond to different convex quadrilaterals whose vertices have different adjacent relations. Clearly the intersection $\vec{p}$ of two diagonals of the quadrilateral together with $\{\vec{p}_k\}^4_{k=1}$ satisfy the equation (\ref{ftp}) while
\be
\frac{\vec{p}-\vec{p}_1}{\|\vec{p}-\vec{p}_1\|}+
\frac{\vec{p}-\vec{p}_4}{\|\vec{p}-\vec{p}_4\|}=\vec{0}, ~~~ \frac{\vec{p}-\vec{p}_2}{\|\vec{p}-\vec{p}_2\|}+
\frac{\vec{p}-\vec{p}_3}{\|\vec{p}-\vec{p}_3\|}=\vec{0}.
\ee
Hence the Fermat-Torricelli point $\vec{p}_F$ of $\{\vec{p}_k\}^4_{k=1}$ is the intersection of two diagonals of the quadrilateral. The sufficient and necessary condition of $\{\vec{m}_i\}^3_{i=1}$ to be jointly measurable coincides with the sufficient and necessary condition of $\vec{m}_1,\vec{m}_2$ to be jointly measurable according to \cite{yuarxiv},
\be
\sum^4_{k=1}\|\vec{p}_k-\vec{p}_F\|=
\|\vec{p}_1-\vec{p}_4\|+
\|\vec{p}_2-\vec{p}_3\|\leq4
\Longleftrightarrow
\|\vec{m}_1+\vec{m}_2\|+\|\vec{m}_{1-2}\|\leq 2.
\ee
Thus to find the optimal $\{\vec{n}_i\}^3_{i=1}$ we may just set $\vec{n}_3=\vec{m}_3$ and find the optimal $\vec{n}_1$ and $\vec{n}_2$. As is shown in Fig. \ref{1111}, if the four vertices $\{\vec{q}_k\}(\{Q_k\}^4_{k=1})$, which lie on the segment going through $\vec{p}_k$ and $\vec{p}_F$ respectively~$(k=1,2,3,4)$, can constitute a convex quadrilateral, it's Fermat-Torricelli point coincides with $\vec{p}_F$. Then these $\{\vec{q}_k\}^4_{k=1}$ satisfying the condition $\expandafter{\romannumeral1})$, $\expandafter{\romannumeral2})$ and $\expandafter{\romannumeral4})$ together are the solution that we are looking for.

By setting $\|\vec{p}_k-\vec{q}_k\|=r~(k=1,2,3,4)$ we obtain from the conditions $\expandafter{\romannumeral2})$ and $\expandafter{\romannumeral4})$,
\be\nonumber
4r+4=\sum^4_{k=1}\|\vec{q}_k-\vec{p}_F\|+
\sum^4_{k=1}\|\vec{p}_k-\vec{q}_k\|
=\|\vec{p}_1-\vec{p}_4\|+
\|\vec{p}_2-\vec{p}_3\|=
2(\|\vec{m}_{12}\|+\|\vec{m}_1-\vec{m}_2\|).
\ee
Hence,
\be
r=\frac{1}{2}(\|\vec{m}_{12}\|+\|\vec{m}_1-\vec{m}_2\|-2).
\ee
In order to calculate $\{\vec{q}_k\}^4_{k=1}$ we assume that $\delta\|\vec{p}_1-\vec{p}_4\|=\|\vec{p}_1-\vec{q}_1\|=r
=\|\vec{p}_2-\vec{q}_2\|=\sigma\|\vec{p}_2-\vec{p}_3\|$. We have
\be\label{delta}
\begin{aligned}
&\delta=\frac{r}{\|\vec{p}_1-\vec{p}_4\|}
=\frac{\|\vec{m}_{12}\|+\|\vec{m}_1-\vec{m}_2\|-2}
{4\|\vec{m}_{12}\|},\\
&\sigma=\frac{r}{\|\vec{p}_2-\vec{p}_3\|}
=\frac{\|\vec{m}_{12}\|+\|\vec{m}_1-\vec{m}_2\|-2}
{4\|\vec{m}_1-\vec{m}_2\|}.\\
\end{aligned}
\ee
With the expressions of $\delta,\sigma$ and $\{\vec{p}_k\}^4_{k=1}$ we obtain the following theorem:

\begin{theorem}[]\label{thm2}The optimal jointly measurable POVMs $\{N^i\}^3_{i=1}$ are given by
\be\label{n123}
\left\{
\begin{aligned}
&\vec{n}_1=
\vec{m}_1-\delta\vec{m}_{12}-\sigma\vec{m}_{1-2},\\
&\vec{n}_2=
\vec{m}_2-\delta\vec{m}_{12}+\sigma\vec{m}_{1-2},\\
&\vec{n}_3=\vec{m}_3,\\
\end{aligned}
\right.
\ee
where $\delta,\sigma$ are given by (\ref{delta}). The corresponding quantum state $\rho$ which gives the optimal value in (\ref{ur3}) is given by the Bloch vectors $\vec{r}_o\in \{\frac{\vec{m}_{12}}{\|\vec{m}_{12}\|}, \frac{\vec{m}_{1-2}}{\|\vec{m}_{1-2}\|}\}$.
\end{theorem}

{\sf [Proof]} According to the analysis above and the locations of $\{\vec{q}_k\}^4_{k=1}$ shown in Fig. \ref{1111}, we have $\vec{p}_k-\vec{q}_k=\delta(\vec{p}_k-\vec{p}_s)~(k\neq s \in\{1,4\})$ and $\vec{p}_l-\vec{q}_l=\sigma(\vec{p}_l-\vec{p}_t)~(l\neq t\in\{2,3\})$. This leads to
\be
\begin{aligned}
&\vec{q}_1=\vec{p}_1-\delta(\vec{p}_1-\vec{p}_4),\\
&\vec{q}_2=\vec{p}_2-\sigma(\vec{p}_2-\vec{p}_3),\\
&\vec{q}_3=\vec{p}_3-\sigma(\vec{p}_3-\vec{p}_2),\\
&\vec{q}_4=\vec{p}_4-\delta(\vec{p}_4-\vec{p}_1).\\
\end{aligned}
\ee

Thus the optimal $\{\vec{n}_i\}^3_{i=1}$ have form
\be\nonumber
\left\{
\begin{aligned}
&\vec{n}_1=\frac{1}{2}(\vec{q}_1+\vec{q}_2)=
\vec{m}_1-\delta\vec{m}_{12}-\sigma\vec{m}_{1-2},\\
&\vec{n}_2=\frac{1}{2}(\vec{q}_1+\vec{q}_3)=
\vec{m}_2-\delta\vec{m}_{12}+\sigma\vec{m}_{1-2},\\
&\vec{n}_3=\frac{1}{2}(\vec{q}_1+\vec{q}_4)=
\vec{m}_3,\\
\end{aligned}
\right.
\ee
where $\delta$ and $\sigma$ are given by (\ref{delta}).
And the optimal state $\rho$ has Bloch vector
$$
\begin{aligned}
&\vec{r}_0=\frac{\vec{p}_1-\vec{q}_1}
{\|\vec{p}_1-\vec{q}_1\|}=\frac{\vec{p}_1-\vec{q}_4}
{\|\vec{p}_1-\vec{q}_4\|}=\frac{\vec{m}_{12}}
{\|\vec{2}_{12}\|},\\
&or\\
&\vec{r}_0=\frac{\vec{p}_2-\vec{q}_2}
{\|\vec{p}_2-\vec{q}_2\|}=\frac{\vec{p}_2-\vec{q}_3}
{\|\vec{p}_2-\vec{q}_3\|}=\frac{\vec{m}_{1-2}}
{\|\vec{m}_{1-2}\|},\\
\end{aligned}
$$\qed

Actually in \cite{wenma} the optimal jointly measurable POVMs approximating two incompatible qubit observables $\vec{m}_1$ and $\vec{m}_2$ are given by
\be\label{n12}
\left\{
\begin{aligned}
&\vec{n}_1=\frac{1}{2}\Big[
(1+\frac{\|\vec{m}_{12}\|-\|\vec{m}_{1-2}\|}{2})\frac{\vec{m}_{12}}{\|\vec{m}_{12}\|}
+(1+\frac{\|\vec{m}_{1-2}\|-\|\vec{m}_{12}\|}{2})\frac{\vec{m}_{1-2}}{\|\vec{m}_{1-2}\|}\Big],\\
&\vec{n}_2=\frac{1}{2}\Big[
(1+\frac{\|\vec{m}_{12}\|-\|\vec{m}_{1-2}\|}{2})\frac{\vec{m}_{12}}{\|\vec{m}_{12}\|}
-(1+\frac{\|\vec{m}_{1-2}\|-\|\vec{m}_{12}\|}{2})\frac{\vec{m}_{1-2}}{\|\vec{m}_{1-2}\|}\Big].
\end{aligned}
\right.
\ee
Substituting $\delta=
\frac{(\|\vec{m}_{12}\|+\|\vec{m}_{1-2}\|)-2}{4\|\vec{m}_{12}\|}$ and $\sigma=
\frac{(\|\vec{m}_{12}\|+\|\vec{m}_{1-2}\|)-2}{4\|\vec{m}_{1-2}\|}$ into (\ref{n123}), one immediately  obtains (\ref{n12}), namely, (\ref{n123}) and (\ref{n12}) coincide.

\begin{figure}[H]
\setlength{\abovecaptionskip}{0.cm}
\setlength{\belowcaptionskip}{-0.cm}
  \centering
  \includegraphics[width=0.38\textwidth,height=4cm]{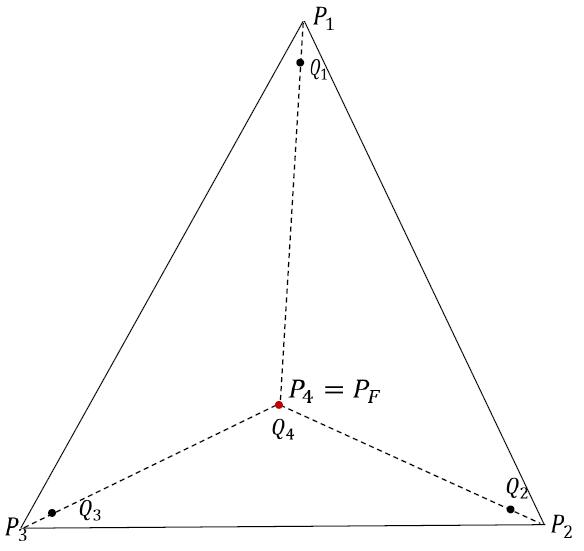}\\
  \caption{Three vertices $(P_1,P_2,P_3)$ of $\{\vec{p}_k\}^4_{k=1}$
   constitute a triangle. $P_4$ lies in the enclosed area of the triangle.
   The Fermat-Torricelli point $\vec{p}_F=\vec{p}_4$.
   There exists no $\{\vec{q}_k\}^4_{k=1}~$ satisfying the conditions
   $\expandafter{\romannumeral1})-
  \expandafter{\romannumeral4})$ simultaneously. %$\expandafter{\romannumeral1}).~
  %\vec{p}_k,\vec{q}_k,\vec{p}_F~(k=1,2,3,4)$ are collinear;
  %$\expandafter{\romannumeral2}).~
  % |\vec{p}_k-\vec{q}_k|=
  %|\vec{p}_l-\vec{q}_l|~(k\neq l;k,l=1,2,3,4)$;
  %$\expandafter{\romannumeral3}). ~
  %\vec{p}_F=\vec{q}_F$;
  %$\expandafter{\romannumeral4})\sum^4_{k=1}|\vec{q}_k-\vec{p}_F|=
  %4$.
  }\label{222}
\end{figure}

Second we consider the situation when $\mid k_1\mid+\mid k_2\mid\geq1$. In this case there are three points with respect to three of the four vectors $\{\vec{p}_k\}^4_{k=1}$ which constitute a triangle. The fourth point with respect to the left fourth vector lies in the enclosed area of the triangle when $\mid k_1\mid+\mid k_2\mid\geq1$, and on the boundary of the area when $\mid k_1\mid+\mid k_2\mid=1$, see Fig. \ref{222}.
The Fermat-Torricelli point $\vec{p}_F=\vec{p}_l$ if $\|\sum_{k\neq l}\frac{\vec{p}_k-\vec{p}_l}
{\|\vec{p}_k-\vec{p}_l\|}\|\leq1$ for some $l\in\{1,2,3,4\}$ \cite{boltyanski}. We just need to analyze the case when $k_1,k_2\geq0$ and $k_1+k_2\geq1$. Since the sum of three distinct unit vectors, which have the same starting points and the sum of interior angles $2\pi$ has norm no more than 1, implies that $\|\sum^3_{k=1}\frac{\vec{p}_k-\vec{p}_4}
{\|\vec{p}_k-\vec{p}_4\|}\|\leq1$, we have $\vec{p}_F=\vec{p}_4$ in this case. It is obvious from Fig. \ref{222} that the condition $\romannumeral2)$ and the condition $\romannumeral3)$ cannot be satisfied simultaneously by any $\{\vec{q}_k\}^4_{k=1}$.

Since no $\{\vec{q}_k\}^4_{k=1}$ can meet the conditions which ensure the optimal approximation, the statistical distance between the POVMs $\{M^i\}^3_{i=1}$ and $\{N^i\}^3_{i=1}$ is strictly larger than the lower bound of (\ref{ur3}), that is,
$$\sum^3_{i=1}\Delta(M^i,N^i)^2>
\frac{1}{2}[\sum^4_{k=1}\|\vec{p}_k-\vec{p}_F\|-4].
$$
Nevertheless, if we set $\vec{q}_4=\vec{q}_F=\vec{p}_F=\vec{p}_4$; $\vec{q}_k$, $\vec{p}_k$ $(k=1,2,3)$ and $\vec{p}_F$ are collinear; $\|\vec{q}_k-\vec{p}_k\|=\|\vec{q}_l-\vec{p}_l\|$ $(k\neq l,~k,l=1,2,3)$ and $\sum^3_{k=1}\|\vec{q}_k-\vec{q}_F\|=4$ (see $\{Q_k\}^4_{k=1}$ in Fig. \ref{222}), we can get that
\be
\sum^3_{i=1}\Delta_{\rho}(M^i,N^i)^2=
\frac{2}{3}[\sum^4_{k=1}\|\vec{p}_k-\vec{p}_F\|-4],
\ee
where $\rho=\frac{1}{2}(I+\vec{r}\cdot\vec{\sigma})$ with $\vec{r}=\frac{\vec{p}_k-\vec{p}_4}
{\|\vec{p}_k-\vec{p}_4\|} ~(k=1,2,3)$, i.e. $\vec{r}\in\{\frac{\vec{m}_{12}}{\|\vec{m}_{12}\|},
\frac{\vec{m}_{1}-\vec{m}_{3}}{\|\vec{m}_{1}-\vec{m}_{3}\|},
\frac{\vec{m}_{2}-\vec{m}_{3}}{\|\vec{m}_{2}-\vec{m}_{3}\|}\}$.

\section{Conclusion}\label{sec3}

Uncertainty principle is one of the distinguished features of quantum physics. For triple incompatible unbiased qubit POVM measurements, we have presented the analytical expressions of the optimal approximations of the jointly measurable POVMs and the optimal measured states, based on the lower bound of the uncertainty relations (\ref{ur3}).
The results on two POVMs \cite{busch14pra,yuarxiv} have spurred the corresponding experimental investigations \cite{wenma,maoyali,kamran}. Our conclusions may result in direct experimental verifications of the uncertainty relations with respect to three POVMs in the optimized schemes.

%%=============================================%%
%% For presentation purpose, we have included  %%
%% \bigskip command. please ignore this.       %%
%%=============================================%%

\backmatter

%\bmhead{Supplementary information}

\bmhead{Acknowledgments}
\bigskip
This work is supported by the NSF of China under Grant Nos. 11701128, 12075159 and 12171044; Beijing Natural Science Foundation (Grant No. Z190005); Academy for Multidisciplinary Studies, Capital Normal University; Shenzhen Institute for Quantum Science and Engineering, Southern University of Science and Technology (No. SIQSE202001); Academician Innovation Platform of Hainan Province.

\bmhead{Data availability}
All the data that support the findings of this work
is accessible in the main manuscript.

%\section*{Declarations}





%%\end{appendices}


\bibliography{sn-bibliography}
\bibliographystyle{splncs03}

\bibliography{referencesTest}

\end{document}